\documentstyle[12pt]{article}
\makeatletter
\newcommand{\singlespacing}{\let\CS=\@currsize\renewcommand{\baselinestretch}
{1.0}\tiny\CS}
\newcommand{\doublespacing}{\let\CS=\@currsize\renewcommand{\baselinestretch}
{1.5}\tiny\CS}
\begin{document}
\textwidth 16cm
\newcommand{\bd}{\begin{document}}
\newcommand{\ed}{\end{document}}
\newcommand{\bc}{\begin{center}}
\newcommand{\ec}{\end{center}}
\newcommand{\bfr}{\begin{flushright}}
\newcommand{\efr}{\end{flushright}}
\newcommand{\lt}{\left}
\newcommand{\rt}{\right}
\newcommand{\vs}{\vspace}
\newcommand{\hs}{\hspace}
\newcommand{\beq}{\begin{equation}}
\newcommand{\eeq}{\end{equation}}
\newcommand{\lb}{\linebreak}
\newcommand{\pb}{\pagebreak}
\newcommand{\mb}{\makebox}
\newcommand{\fb}{\framebox}
\newcommand{\mc}{\multicolumn}
\newcommand{\ben}{\begin{enumerate}}
\newcommand{\een}{\end{enumerate}}
\newcommand{\bit}{\begin{itemize}}
\newcommand{\eit}{\end{itemize}}
\newcommand{\ol}{\overline}
\newcommand{\un}{\underline}
\newcommand{\lefq}{\lefteqn}
\newcommand{\ba}{\begin{array}}
\newcommand{\ea}{\end{array}}
\newcommand{\beqa}{\begin{eqnarray}}
\newcommand{\eeqa}{\end{eqnarray}}
\newcommand{\beqas}{\begin{eqnarray*}}
\newcommand{\eeqas}{\end{eqnarray*}}
\newcommand{\bfg}{\begin{figure}}
\newcommand{\efg}{\end{figure}}
\newcommand{\bds}{\begin{displaymath}}
\newcommand{\eds}{\end{displaymath}}
\newcommand{\btb}{\begin{tabbing}}
\newcommand{\etb}{\end{tabbing}}
\newcommand{\para}{\parallel}
\newcommand{\pad}{\partial}
\newcommand{\nn}{\nonumber}
\newcommand{\la}{\leftarrow}
\newcommand{\ra}{\rightarrow}
\newcommand{\lgla}{\longleftarrow}
\newcommand{\lgra}{\longrightarrow}
\newcommand{\La}{\Leftarrow}
\newcommand{\Ra}{\Rightarrow}
\newcommand{\Lra}{\Leftrightarrow}
\newcommand{\Lgla}{\Longleftarrow}
\newcommand{\Lgra}{\Longrightarrow}
\newcommand{\bm}{\boldmath}
\newcommand{\lan}{\langle}
\newcommand{\ran}{\rangle}
\renewcommand{\a}{\alpha}
\renewcommand{\b}{\beta}
\newcommand{\g}{\gamma}
\newcommand{\G}{\Gamma}
\renewcommand{\d}{\delta}
\newcommand{\eps}{\epsilon}
\newcommand{\th}{\theta}
\newcommand{\Th}{\Theta}
\newcommand{\s}{\sigma}
\newcommand{\lam}{\lambda}
\newcommand{\D}{\Delta}
\newcommand{\vare}{\varepsilon}
\newcommand{\pr}{\prime}
\newcommand{\ro}{\rho}
\newcommand{\nab}{\nabla}
\newcommand{\m}{\mu}
\newcommand{\n}{\nu}
\newcommand{\Sg}{\Sigma}
\newcommand{\p}{\pi}
\newcommand{\R}{I\!\!R}
\newcommand{\om}{\omega}
\newcommand{\Om}{\Omega}
\newcommand{\ze}{\zeta}
\newcommand{\vart}{\vartheta}
\newcommand{\tri}{\triangle}
\newcommand{\f}{\frac}
\newcommand{\iny}{\infty}
\newcommand{\pro}{\propto}
%\input{fqhsc}
%\input{state}
%\input{acc}
%\input{fs}
%\ed

\bc

{\large \bf ISOSPECTRAL PARTNERS OF A COMPLEX } \\
{\large \bf ${ \cal{PT}}$-INVARIANT POTENTIAL }

\ec

\vs{1cm}

\bc
{\large \bf Anjana Sinha $^*$} \\
{\it Dept. of Applied Mathematics}\\
{\it University of Calcutta}\\
{\it 92 A.P.C.Road, Kolkata - 700 009}\\

\vs{.5cm}

{\it and} \\

\vs{.5cm}

{\large \bf Rajkumar Roychoudhury $^{\$}$}\\
{\it Physics \& Applied Mathematics Unit}\\
{\it Indian Statistical Institute}\\
{\it Kolkata - 700 108}

\ec

%\date{}

\vs{3cm}

\noindent
-----------------------------------------------------------
------------------------------------- \\
e-mail : \\
$^*$ a.sinha@cucc.ernet.in, anjana23@rediffmail.com \\
$^{\$}$ raj@isical.ac.in \\

\pb

\bc
{\bf \large \un{Abstract}}
\ec

We construct isospectral partner potentials of a complex ${\cal{PT}}$-invariant
potential, viz., $V(x) = -V_1 ~sech^2 ~x ~-~ i V_2 ~sech ~x ~~tanh ~x $ 
using Darboux's method. One set of isospectral potentials are obtained which
can be termed 'Satellite potentials', in the sense that they are of the same
form as the original potential. In a particular case, the supersymmetric
partner potential has the same spectrum, including the zero energy
ground state, a
fact which cannot occur in conventional supersymmetric quantum mechanics with
real potential. An explicit example of a non-trivial set of 
isospectral potential is also obtained. 

\vspace{5cm}

\noindent
-----------------------------------------------------------
------------------------------------- \\

\noindent
{\bf \un{Key words :}} $\cal{PT}$- symmetry, non-Hermitian Hamiltonian, 
isospectral potential, Darboux's method, supersymmetry.

\newpage

\section*{I. Introduction}

Quantum systems characterized by non-Hermitian Hamiltonians have been studied
from time to time, because of their applications in scattering problems [1].
For this reason, complex potentials were also termed as {\it optical} or 
{\it average nuclear potentials}. The interest in this field has been revived
fairly recently after it was conjectured by Besis, Bender and Boettcher, 
and others, that Hermiticity of the Hamiltonian is not essential for
a real spectrum [2,3]. Several non-Hermitian ${ \cal{PT}}$-invariant complex
potentials have been found to possess a real spectrum, though the
corresponding Hamiltonians are non-Hermitian [4,5]. Moreover, ${
\cal{PT}}$-invariant models admit some of the properties of the usual
Hermitian ones, {\it viz.}, supersymmetry, potential algebra,
quasi-solvability, etc. [6-17]. Such non-Hermitian ${ \cal{PT}}$-invariant
Hamiltonians have found applications in many areas of theoretical physics --- 
nuclear physics, field theories when studying Lee-Yang zeros,
localization-delocalization transitions in superconductors, theoretical
description of defraction of atoms by standing light waves, and also the
study of solitons on a complex Toda lattice [7].

In this study, our aim is to construct new complex potentials via the
Darboux's method [7-10], for which the
corresponding eigenvalue problem can be solved exactly. The constructed
potentials are not necessarily ${ \cal{PT}}$-invariant, but still give rise to
a real and discrete spectrum, provided the original potential admits real
energies only.

The organization of the present note is as follows. To make it self-contained,
we give a brief review of ${ \cal{PT}}$-symmetry in Section II. 
In Section III, we give an outline of the Darboux's method for constructing
isospectral partner potentials of a given known ${ \cal{PT}}$-symmetric 
potential. 
In Section IV, we illustrate our approach with the help of 
an explicit example, {\it viz.}, the one-dimensional ${ \cal{PT}}$-invariant
potential 
\beq
V(x) = - V_1 ~sech ^2 ~x - i V_2 ~sech ~x ~~ tanh ~x \ \ \ \ \ \ \ \ V_1 > 0 
\eeq
Section V is kept for conclusions and discussions.

\vs{2cm}

\section*{II. ${ \cal{PT}}$-symmetry}

The Hamiltonian ${\cal{H}}$ for a particle of mass $m$, in a complex 
potential \\
$ V(x) = V_R (x) + i V_I (x) $ is given by
\beq
{ \cal{H}} ~=~ - \f{1}{2m} \f{d^2}{dx^2} ~+~ V(x) 
\eeq
${ \cal{H}}$ is said to be ${ \cal{PT}}$-symmetric when
\beq
{ \cal{PTH}} ~=~ { \cal{HPT}}
\eeq
Here ${ \cal{P}}$ is the {\it Parity operator} acting as spatial reflection,
and ${ \cal{T}}$ stands for {\it Time Reversal}, acting as the complex
conjugation operator. Their action on the position and momentum operators are
given by :
$${ \cal{P}} ~:~ x \ra -x, ~~~p \ra -p $$
$${ \cal{T}} ~:~ x \ra x, ~~~~p \ra -p, ~~~~ i \ra -i $$ 
Hence, in explicit form, the condition for a potential to be ${
\cal{PT}}$-symmetric is 
\beq
V^* (-x) ~=~ V(x)
\eeq
The commutation relation 
\beq
\lt [ x, p \rt ] = i \hbar 
\eeq
remains invariant under ${ \cal{PT}}$ for both real as well as complex $x$ and
$p$. 

It is worth mentioning here that though the Hermiticity of the Hamiltonian may
be replaced by the weaker condition of ${ \cal{PT}}$-symmetry, the latter is
not sufficient for the reality of the spectrum. Various authors have studied
several one-dimensional non-Hermitian ${ \cal{PT}}$-symmetric models and shown
that such Hamiltonians exhibit 2 types of behaviour --- \\
(i) ~~ In the unbroken ${ \cal{PT}}$-symmetry phase, the eigenfunctions of 
${\cal{H}}$ are also eigenfunctions of ${ \cal{PT}}$, and the energy spectrum
is real and discrete. \\
(ii) ~~ In spontaneous breakdown of ${ \cal{PT}}$-symmetry, though the
potential retains ${ \cal{PT}}$-symmetry, the corresponding wavefunctions do
not, and the energy eigenvalues exist as complex conjugate pairs.

\section*{III. Darboux's method}

The Darboux's method [7-10] relates the spectral properties of a pair of 
standard Schr\"{o}dinger Hamiltonians
\beq
H_{\pm} = - \f{\hbar ^2}{2m} \f{d^2}{dx^2} + V_{\pm} (x) 
\eeq
We assume that the spectral properties of one of these Hamiltonians, say 
$H_{+}$ , is exactly known. Thus
\beq
H_{+} \phi _n (x) = E_n \phi _n (x) 
\eeq
where the eigenvalues $E_n$ and the corresponding eigenfunctions 
$ \phi _n (x) $ are known explicitly. The spectrum is assumed to be discrete
such that $ E_0 < E_1 < E_2 < \cdots $. \\
Let there exist a linear operator $A$, such that it obeys an intertwining
relationship 
\beq
A H_{+} = H_{-} A
\eeq
Hence the functions $  \psi _n = A \phi _n $ are eigenfunctions of $H_{-}$ 
with the same eigenvalues $E_n$
\beq
H_{-} \psi _n (x) = E_n \psi _n (x)
\eeq
A general form for an intertwining operator $A$ obeying (3) may be 
given by [7,8]
\beq
A = \sum _{k=0} ^N f_k (x) \f{d^k}{dx ^k}
\eeq
where $f_k (k=0, 1, 2, ...... N-1 )$are at least twice differential functions
and $f_N$ is an arbitrary constant.\\
For simplicity of calculations, we work in units $\hbar = 2m = 1$.

We take the simplest non-trivial choice for $A (N=1)$
\beq
A = - \f{d}{dx} + f(x)
\eeq
Putting (6) in (3) yields
\beq
\lt [ V_{-} - V_{+} + 2 f^{\pr} \rt ] \f{d}{dx} 
- \lt [ \lt ( V_{-} - V_{+} \rt ) f + V_{+} ^{\pr} - f^{~\pr \pr} \rt ].1
= 0 
\eeq
where prime denotes differentiation with respect to $x$. This relationship 
(12) puts certain restrictions on the functions $V_{+} (x), V_{-} (x), $ and 
$f(x)$, viz.,
\beq
V_{-} (x) = V_{+} (x) - 2 f^{~\pr} (x) 
\eeq
\beq
\lt [ V_{-} (x) - V_{+} (x) \rt ] f(x) + V_{+} ^{\pr} (x) -  
f^{~\pr \pr} (x) = 0
\eeq
Substituting (14) in (13), and integrating, we obtain
\beq
f^{~2} (x) + f^{~ \pr} (x) - V_{+} (x) = - \eps
\eeq
where $ \eps $ is an arbitrary integration constant, in general complex.\\
Putting
\beq
f(x) = \f{u ^{~ \pr} (x) }{u(x)}
\eeq
(15) can be cast in the form of Schr\"{o}dinger-like equation
\beq
- \f{d^2 u(x)}{dx^2} + V_{+} u(x) = \eps u(x) 
\eeq
$ \eps $ is sometimes called the {\it factorization energy}. It is worth
noting here that $u(x)$ need not be square-integrable. So we are not
restricted to normalizable solutions of (17). However, for $A$ to be well
defined, $u$ must not have any zeroes on the real line. For this,
it requires the condition $ \eps < E_0 $. 

In terms of the function $f(x)$, the two potentials are expressed as
\beq
V_{\pm} (x) = f^{~2} (x) \pm f^{~ \pr} (x) + \eps
\eeq
At this point it is obvious that for $ \eps = 0 $, the two potentials 
$ V_{\pm}(x) $ are the supersymmetric (SUSY) partner potentials, and 
$H_{\pm}$ are the SUSY partner Hamiltonians. From (18), the partner
Hamiltonians may be expressed in terms of the linear operator $A$ as,
\beq
H_{+} = A^{+} A + \eps
\eeq
\beq
H_{-} = AA^{+}  + \eps
\eeq
We shall apply the above approach to construct a new complex potential 
$V_{-}(x)$, which is isospectral with a known $\cal{PT}$-symmetric potential 
$V_{+}(x)$ with real eigenvalues, obeying the relationship
\beq
V_{-}(x) = 2 \lt \{ \f{u{\pr}(x)}{u(x)} \rt \}^2 - V_{+}(x) + 2 \eps
\eeq
Though ${\cal{PT}}$-symmetry may be either restored or broken in the 
new potential, the partner Hamiltonian $H_{-}$ has the same eigenvalues as 
$H_{+}$ (with the possible exception of the ground state). The
corresponding eigenfunctions are given by
\beq
\psi _n (x) = c_n \lt \{ - \f{d \phi _n (x)}{dx} +   \f{u{\pr}(x)}{u(x)}
\phi _n (x) \rt \}
\eeq
Here $c_n$ is the normalization constant, given by
\beq
\begin{array}{lcl}
\lt| c_n \rt|^{-2} &=& \lt< A \phi _n \vert A \phi _n \rt> \\
&=& E_n - \eps\\
\end{array}
\eeq
We shall make our ideas clear with the help of an explicit 
example in the next section.

\section*{IV. Explicit example}

As an explicit example, we shall consider the one-dimensional
$\cal{PT}$-invariant potential [10,18]
\beq
V(x) = - V_1 ~sech ^2 ~x - i V_2 ~sech ~x ~~ tanh ~x \ \ \ \ \ \ \ \ V_1 > 0 
\eeq
This non-Hermitian Hamiltonian has the interesting property of having both
real and complex discrete spectrum, depending on the relative magnitudes of
its real and imaginary parts, viz., $V_1$ and $V_2$. \\
(i) ~~~ For $ \lt | V_2 \rt | \leq V_1 + \f{1}{4} $ \\
$\cal{PT}$-symmetry is unbroken and energies are real. \\
(ii) ~~ For $ \lt | V_2 \rt | > V_1 + \f{1}{4} $ \\
$\cal{PT}$-symmetry is spontaneously broken and energies are complex conjugate
pairs. \\
We shall restrict our attention to the unbroken $\cal{PT}$-symmetric phase,
with real energies only.

To solve the Schr\"{o}dinger-like equation
$$ - \f{d^2 u(x)}{dx^2} + V(x) u(x) = \eps u(x) $$
with $V(x)$ given by (24), we make the following substitutions :
\beq
z = \f{1 - i ~sinh ~x}{2}
\eeq
\beq
u(x) = z^{-p} ~(1-z)^{-q} ~ \chi (z)
\eeq
Then the differential equation satisfied by $ \chi (z) $ is
\beq
z (1-z) \chi ^{~\pr \pr} + \lt[ -2p + \f{1}{2} + z \lt( 2p + 2q - 1 \rt)
\rt] \chi ^{~ \pr} - \lt[ \lt( p + q \rt) ^2 - \lam ^2 \rt] \chi = 0
\eeq
provided
\beq
p = - \f{1}{4} \pm \f{s}{2}
\eeq
\beq
q = - \f{1}{4} \pm \f{t}{2}
\eeq
\beq
t = \sqrt{ V_1 - V_2 + \f{1}{4} }
\eeq
\beq
s = \sqrt{ V_1 + V_2 + \f{1}{4} }
\eeq
\beq
\lam ^2 = - \eps
\eeq
The most general solution of (27) is 
\beq
\chi (z) = \alpha ~ F \lt( a, b, c, z \rt) + \beta z^{1-c} ~ (1-z) ^{c-a-b}
~ F \lt( 1-a, 1-b, 2-c, z \rt)
\eeq
where
\beq
a = - p - q + \lam
\eeq
\beq
b = - p - q - \lam
\eeq
\beq
c = - 2p  + \f{1}{2}
\eeq
So the general form of $u$ is
\beq
u = \alpha ~ z^{-p} ~ (1-z) ^{-q} ~ F \lt( a, b, c, z \rt) 
+ \beta z^{p + 1/2} ~ (1-z) ^{q + 1/2} ~ F \lt( 1-a, 1-b, 2-c, z \rt)
\eeq
Since $u$ must not have any real zero, so $ \alpha$ must be non-zero. This
allows us to put $ \alpha = 1 $. However, the proper choice of the sign in the
expressions for $p$ and $q$ is extremely important.

For the potential given in (24), the real energy eigenvalues and the
corresponding eigenfunctions are explicitly given by [18],
\beq
E_n = - \lt \{ n + \f{1}{2} - \f{1}{2} \lt( s + t \rt) \rt \} ^2 
\ \ \ \ \ n = 0, 1, 2, .... < \f{s+t-1}{2}
\eeq
\beq
\phi _n (x) = \lt( \begin{array}{c} n\\ n-2p-\f{1}{2} \\  \end{array} \rt)
\lt( \f{1-i ~ sinh ~x}{2} \rt) ^{-p} \lt( \f{1+i ~ sinh ~x}{2} \rt) ^{-q}
P_n ^{-2p -\f{1}{2}, -2q -\f{1}{2}} (i ~sinh ~x)
\eeq
where $P_n ^{a,b} (z) $ are the Jacobi polynomials expressed as [19]
\beq
P_n ^{a,b} (z) = 2 ^{-n} \sum _{m=0}^n \lt( \begin{array}{c} n+a\\ m\\ 
 \end{array} \rt) ~\lt( \begin{array}{c} n+b\\ n-m\\ \end{array} \rt) 
(z-1)^{n-m} (z+1)^m 
\eeq

\vspace{1cm}

\noindent
{\bf \un{Case (i)}} 

First let us consider the simplest case given by
$$ \beta = 0 $$
$$ b = c $$
Hence 
\beq
\lam = p - q - \f{1}{2}
\eeq
\beq 
\eps = - \lam ^2 = - \lt[ p - q - 1/2 \rt] ^2
\eeq
Using the relationship [20]
\beq
F(a, b, b, z) = (1-z)^{-a}
\eeq
$u(x)$ reduces to the simple form
\beq
u(x) = \alpha \lt( \f{1-i~sinh ~x}{2} \rt) ^{-p}
\lt( \f{1+i~sinh ~x}{2} \rt) ^{q+1/2}
\eeq
Thus $f(x) = u^{\pr}(x)/u(x) $ is obtained to be
\beq
f(x) = i ~\lt[ p + \lt ( q + \f{1}{2} \rt) \rt] sech ~x ~-~ 
\lt[ p - \lt ( q + \f{1}{2} \rt) \rt] tanh ~x 
\eeq
The isospectral partner potential of (24) as obtained from (21) turns out 
to be
$$ V_{-} (x) ~=~ - \lt[ 4 \lt \{ p^2 + \lt( q + \f{1}{2} \rt) ^2 \rt \} 
- V_{1} \rt] sech ^2 x $$
\beq
~~~~~~~~~~~~~~ ~-~ i~ \lt[ 4 \lt \{ p^2 - \lt( q + \f{1}{2} \rt) 
^2 \rt \} - V_{2} \rt] sech ~x ~~tanh ~x 
\eeq
with the corresponding eigenfunction (from (22)) $\psi _n$ as 
$$ \psi _n = \lt[ \lt( \f{1-i ~ sinh ~x}{2} \rt) ^{-p} 
\lt( \f{1+i ~ sinh ~x}{2} \rt) ^{-q} \rt] $$
$$ \lt[ \lt( \begin{array}{c} n\\ n-2p- \f{1}{2} \\  \end{array} \rt)
\lt( 2q + \f{1}{2} \rt) \lt[ i ~sech ~x + tanh ~x \rt] 
P_n ^{-2p - \f{1}{2}, -2q - \f{1}{2} } (i ~sinh ~x) \rt. $$
\beq
\lt. ~-~ \f{(p+q)^2-E_n}{-2p+1/2}
\lt( \begin{array}{c} n-1 \\ n-2p- \f{1}{2} \\  \end{array} \rt)
\lt( 2q + \f{1}{2} \rt) \lt[ i ~sech ~x + tanh ~x \rt] 
P_{n-1} ^{-2p + \f{1}{2}, -2q + \f{1}{2} } (i ~sinh ~x) \rt ]
\eeq

This is the special case of the so-called satellite potentials, with 
$V_{+}(x)$ and $V_{-}(x)$ having exactly the same form, but with different
coefficients. \\
We illustrate this with the help of some simple  examples. \\
Let  $ \ \ \ \ \ V_1 = 25, \ \ \ \ \ V_2 = 5 $ \\
Thus 
\beq
V_{+} (x) = -25 ~sech ^2 x - 5i ~sech ~x ~~tanh ~x
\eeq
with energies
\beq
E_n = - \lt \{ n + \f{1}{2} - 5 \rt \} ^2 , \ \ \ \ n = 0, 1, 2, 3, 4 
\eeq
Then using equations $(28)$ - $(31)$, the values of $p$ and $q$ are \\
$ p = \f{5}{2} $ ~~~~~~~ or ~~~~~~~ $ - 3 $ \\
$ q = 2 $ ~~~~~~~ or ~~~~~~~ $ - \f{5}{2} $ \\
In order that the wave function vanishes asymptotically, only the positive sign
of the discriminant is allowed for $p$. Therefore, $ p = 5 / 2 $.
The two values of $q$ give rise to two interesting cases. 

\vs{.5cm}

\noindent
{\bf a) ~~ $ q = - 5 / 2  $ \\}
$ f(x) $ for this particular case becomes 
\beq
f(x) =  \f{11}{2}  tanh ~x ~-~  \f{i}{2}  sech ~x  
\eeq
and the isospectral partner potential of (48) turns out to be
\beq
V_{-} (x) = -16 ~sech ^2 x - 4i ~sech ~x ~~tanh ~x
\eeq
with energies
\beq
E_n = - \lt \{ n + \f{1}{2} - 4 \rt \} ^2 , \ \ \ \ n = 0, 1, 2, 3, 
\eeq
It is easy to check that $ V_{\pm}(x)$ share exactly
the same energy spectrum with the 
exception of the ground state, as shown in Table 1 below.

\begin{center}

{\bf Table 1 } 

\vs{.25cm}

\begin{tabular}{|c|c|c|} \hline
$n$ & $ E_n [ V_+(x) ]$ & $ E_n [ V_-(x) ] $ \\ \hline
$0$ & $ - 81 / 4 $ & $ - 49 / 4 $ \\ \hline
$1$ & $ - 49 / 4 $ & $ - 25 / 4  $ \\ \hline
$2$ & $ - 25 / 4 $ & $ - 9 / 4 $ \\ \hline
$3$ & $ - 9 / 4 $ & $ - 1 / 4 $ \\ \hline
$4$ & $ - 1 / 4 $  & ---  \\ \hline
\end{tabular}
\end{center}

\noindent
{\bf b) ~~ $ q = 2 $ \\}
This is the special case when
$ p = q + 1 / 2 $~~, ~~so that $ \eps = 0 $ \\
and $ f(x) $ becomes purely imaginary.
\beq
f(x) = i ~ 5 ~ sech ~x 
\eeq
The isospectral partner reduces to
\beq
V_{-} (x) = -25 ~sech ^2 x ~+~  5~i ~sech ~x ~~tanh ~x
\eeq
Thus 
\beq
V_{\pm}(x) = f^2 (x) \pm f^{\pr} (x)
\eeq
The partner potentials (48) and (54) are totally  degenerate. They share 
identical energies,  including the zero energy 
ground state, and supersymmetry (SUSY) is broken. This scenario is 
quite different from conventional SUSY breaking. In the conventional
supersymmetric quantum mechanics (SUSYQM), SUSY is broken when the zero state
energy does not exist and both $ V_{\pm} (x) $ have the same spectrum. 
The ground state wavefunctions $ \phi _0 (x) $ and $ \psi _0 (x) $ of the
partner potentials (48) and (54), are respectively given by
\begin{equation}
\begin{array}{lcl}
\phi _0 (x) &=&\displaystyle \lt ( \f{1 - i ~sinh ~x }{2} \rt ) ^{-\f{5}{2}} 
 \lt ( \f{1 + i ~sinh ~x }{2} \rt ) ^{-2}\\
&=&\displaystyle sech ^4 x \lt [ sech ~x \sqrt{ 1 + cosh ~x } 
+ i \f{tanh ~x}{\sqrt{1 + cosh ~x}} \rt]\\ \end{array}
\eeq
\beq
\begin{array}{lcl}
 \psi _0 (x) &=&\displaystyle \lt [ - \f{d \phi_0 }{dx} + \f{u ^{\pr} (x) }{u
(x) } \phi _0 (x) \rt]\\
&=&\displaystyle \f{9}{2} \lt [ tanh ~x + i ~sech ~x \rt ] \phi _0 (x) \\
&=&\displaystyle sech ^5 x \lt [ tanh ~x \lt ( \sqrt{ 1 + cosh ~x} 
\f{1}{ \sqrt{ 1 + cosh ~x} } \rt ) \rt. \\
&&\displaystyle +~ \lt. i ~ sech ~x \lt ( \f{sech ^2 x }{ \sqrt{ 1 + cosh ~x}
} +  \sqrt{ 1 + cosh ~x} - \f{1}{ \sqrt{ 1 + cosh ~x}} \rt ) \rt ]\\
\end{array} 
\eeq
apart from normalization constants.

\vs{1cm}

\noindent
{\bf \un{Case (ii)}}

Next we consider the particular case 
$$ \beta = 0, ~~b = a + \f{1}{2}, ~~c = 2a $$
This gives the following values for 
$$ \lam = - \f{1}{4}, ~~\eps = - \f{1}{16}, ~~ q = - \f{1}{2} $$
Hence, for this particular case, $V_1 = V_2 = V_0 $ (say) in (24).
Using the relationship [20]
\beq
F \lt( a, a+\f{1}{2}, 2a, z \rt) = 2^{2a-1}(1-z)^{-1/2} 
\lt[ 1 + (1-z) ^{1/2} \rt] ^{1-2a}
\eeq
$u(x)$ reduces to 
\beq
u(x) = \alpha ~ 2^{-2p-1/2} \lt( \f{1-i~sinh ~x}{2} \rt) ^{-p}
\lt[ 1 + \lt( \f{1+i~sinh ~x}{2} \rt)^{1/2} \rt] ^{2p + 1/2}
\eeq
Thus $f(x) = u^{\pr}(x)/u(x) $ is obtained to be
\beq
f(x) =  \f{1}{4} tanh ~x  ~-~ \f{i}{4}~ sech ~x 
~+~ i ~ \lt ( 2p + \f{1}{2} \rt) sech ~x ~ \lt[ \f{1 + i ~ sinh ~x }{2}
\rt] ^{1/2}
\eeq
The isospectral partner potential of (24) viz.,
\beq
V_{+} (x) = - V_0 ~sech ^2 ~x - i V_0 ~sech ~x ~~ tanh ~x  
\eeq
is obtained from (21), and turns out to be
$$ V_{-} (x) = - \lt[ \f{1}{4} + \lt( 2p + \f{1}{2} \rt) ^2 - V_0 \rt] 
~sech ^2 ~x - i ~ \lt[ \f{1}{4} + \lt(2p + \f{1}{2} \rt) ^2 - V_0 \rt] 
~ sech ~x ~~tanh ~x $$
\beq
~~~~~ + \lt( 2p + \f{1}{2} \rt) \lt[ sech ^2 ~x + i ~ sech ~x ~~ tanh ~x \rt]
\lt( \f{1 + i ~ sinh ~x }{2} \rt) ^{1/2}
\eeq
or more explicitly,
\beq
V_{-} (x) = -V_R (x)~sech^2 x - i V_I (x) ~sech ~x ~~tanh ~x 
\eeq
with 
\beq
V_R (x) =  \sigma + \lt( p + \f{1}{4} \rt) \lt \{ 
\sqrt{ cosh ~x + 1 } + \f{1}{\sqrt{ cosh ~x + 1 }} 
- \f{cosh ^2 x}{ \sqrt{cosh ~x + 1 }} \rt \}
\eeq
\beq
V_I (x) =  \sigma - \lt( p + \f{1}{4} \rt) \lt \{ 
\sqrt{ cosh ~x + 1 } + \f{1}{\sqrt{ cosh ~x + 1 }} \rt \}
\eeq
where
\beq
\sigma = 4 p^2 + 2p + \f{1}{2} - V_0
\eeq
and 
\beq
p = - \f{1}{4} \pm \f{1}{2} \sqrt{ 2 V_0 + \f{1}{4}}
\eeq
It is worth noting here that the coefficients of 
$(sech ^2 x) $ and $ (sech ~x ~~ tanh ~x )$
are pure numbers in the expression for $V_{+}(x)$, 
whereas they are functions of $x$ in $ V_{-}(x)$.
From equation (38) it is essential that $s$ and hence $V_0$ be 
large for sufficient number of energy levels, as $ t = 1 / 4 $. 
This prevents $p$ from taking the value $ - 1 / 4 $.
So the isospectral partners $ V_{\pm}(x)$ in equations 
(61) and (63), cannot be reduced to satellite potentials
for any physical value of $p$ in this particular case.

The corresponding eigenfunction is calculated to be 
$$ \psi _n (x) = 
\lt[ -i ~\lt( p + \f{3}{4} \rt) sech ~x + \lt( p - \f{1}{4} \rt) tanh ~x 
+ i \lt( 2p + \f{1}{2} \rt) sech ~x \lt( \f{1 + i ~sinh ~x }{2} \rt) ^{1/2}
\rt] $$
$$ \lt( \begin{array}{c} n \\ n-2p-1/2 \\  \end{array} \rt)
 \lt( \f{1 - i ~sinh ~x }{2} \rt) ^{-p}
\lt( \f{1 + i ~sinh ~x }{2} \rt) ^{\f{1}{2}}
P_n ^{-2p-\f{1}{2}, \f{1}{2} }(i ~sinh ~x) $$
$$ ~~-~~ \f{ (p-1/2) ^2 -E_n }{-2p + 1/2} 
~\lt( \f{1 - i ~sinh ~x }{2} \rt) ^{-p}
\lt( \f{1 + i ~sinh ~x }{2} \rt) ^{\f{1}{2}} $$
\beq
\lt( \begin{array}{c} n-1 \\ n-2p-1/2 \\  \end{array} \rt)
P_n ^{-2p+\f{1}{2}, \f{3}{2}} (i ~sinh ~x)
\eeq

\section*{V. Conclusions and Discussions}

To conclude, it is shown in this paper how to generate equivalent
SUSY partners of the complex ${\cal{PT}}$- invariant potential,
viz.,  
$$ V(x) = - V_1 sech ^2 x ~-~ i~V_2 sech ~x ~~ tanh ~x $$
with the help of Darboux's method.
The form of the newly constructed potentials  
depend on the choice of $\beta $ and $ \eps $, but
they share the same energy spectrum as the origial potential, with the
possible exception of the ground state.
If the original potential is so chosen that it
admits real eigenvalues only, then we can obtain a series of non-trivial
complex potentials generating the
same real-valued spectrum. The new wavefunctions are also easily obatined 
by this approach. 

Depending on particular values of $V_1$ and $V_2$, the isospectral partners
may be of similar nature with just the coupling constants taking different
values. We term these as {\it satellite potentials}. In the special case 
$ p = 5/2, q = 2, \eps = - (p - q - 1/2 )^2 - 0 $, the
SUSY partner potentials have the same spectrum including the zero energy
ground state, a situation which cannot occur in conventional SUSYQM with real
potential. 

We have also constructed a non-trivial partner potential, and
compared its real and imaginary parts graphically with those of the
original potential, in Figures  1 and 2 respectively. 
It is interesting to note that even in the non-trivial
case, the partner potentials have similar geometric form.

 Further constructions with non-zero
$\beta$ and other values of the integration constant $\eps$ are left as a
future exercise.

\section*{Acknowledgment}

One of the authors (A.S.) is grateful to the Council of Scientific and
Industrial Research, India, for financial assistance.

\pb

\section*{References}

\begin{enumerate}
\item[1.] H. Feshbach, C. E. Porter \& V. F. Weisskopf, Phys. Rev. {\bf 96}
448 (1954).
\item[2.] D. Besis, unpublished (1992).
\item[3.] C. M. Bender \& S. Boettcher, 
Phys. Rev. Lett. {\bf 80} 5243 (1998),
J. Phys. {\bf A 31} L273 (1998).
\item[4.] C. M. Bender, S. Boettcher \& P. N. Meisinger, J. Math. Phys. 
{\bf 40} 2201 (1999).
\item[5.] V. M. Tkachuk \& T. V. Fityo, arXiv : quant-ph/0204018 (2002) 
{\it and references therein}.
\item[6.] B. Bagchi, S. Mallik \& C. Quesne, Int. J. Mod. Phys. {\bf A 17} 
51 (2002) {\it and references therein}.
\item[7.] F. Cannata, G. Junker \& J. Trost, Phys. Lett. {\bf A 246} 219
(1998). 
\item[8.] F. Cannata, G. Junker \& J. Trost, 'Particles, Fields and
Gravitation' , ed. J. Rembielinski, AIP Conf. Proc. 453 (AIP, Woodbury, 1998)
209. 
\item[9.] G. Junker and P. Roy, 
Annals of Phys. {\bf 270} 155 (1998),
Phys. Lett. A {\bf 257} 113 (1999).
\item[10.] B. Bagchi and R. Roychoudhury, J. Phys. A : Math. Gen. {\bf 33}
L1 (2000).
\item[11.] A. A. Andrianov, F. Cannata, J. P. Denonder, M. V. Ioffe, Int. J.
Mod. Phys. A {\bf 14} 2675 (1999).
\item[12.] M. Znojil, F. Cannata, B. Bagchi, R. Roychoudhury, Phys. Lett. B
{\bf 483} 284 (2000).
\item[13.] B. Bagchi, F. Cannata and C. Quesne, Phys. Lett. A {\bf 269} 79
(2000). 
\item[14.] F. Cannata, M. Ioffe, R. Roychoudhury and P. Roy, Phys. Lett. A 
{\bf 281} 305 (2001).
\item[15.] M. Znojil, Phys. Lett. A {\bf 259} 220 (1999), 
Czech. J. Phys. {\bf 51} 420 (2001), 
J. Phys. A {\bf 33} 4561 (2000),
J. Phys. A {\bf 33} L61 (2000),
J. Phys. A {bf 35} 2341 (2002).
\item[16.] P. Dorey, C. Dunning and R. Tateo, J. Phys. A {\bf 34} L391 (2001).
\item[17.] G. L\'{e}vai  and M. Znojil,
J. Phys. A {\bf 33} 7165 (2000),
Mod. Phys. Lett. A {\bf 16} 1973 (2001).
\item[18.] Z. Ahmed, Phys. Lett. {\bf A 282} 343 (2001).
\item[19.] I. S. Gradshteyn \& I. M. Ryzhik, Tables of Integrals, Series and
Products, AP, New York, (1980).
\item[20.] M. Abramowitz \& I. A. Stegun, Handbook of Mathematical Functions,
Dover Pub. Inc., New York, (1970).

\end{enumerate}

\ed